\documentclass[aps,prl,superscriptaddress,showpacs,twocolumn,floatfix]{revtex4-1}
\usepackage{latexsym}
\usepackage{amsthm}
\usepackage{amsmath}
\usepackage{amssymb}
\usepackage{hyperref}
\usepackage{bbm}
\usepackage{color}
\usepackage{colordvi}
\usepackage{comment}
\usepackage{times,latexsym,graphicx,wrapfig,url}
\usepackage{epsfig,lineno,bm}
\usepackage{subfigure}
\usepackage{mciteplus}

\newcommand{\TeV}{\,\mathrm{TeV}}
\newcommand{\GeV}{\,\mathrm{GeV}}
\newcommand{\MeV}{\,\mathrm{MeV}}

\newcommand{\be}{\begin{equation}}
\newcommand{\ee}{\end{equation}}
\newcommand{\bea}{\begin{eqnarray}}
\newcommand{\eea}{\end{eqnarray}}

\newcommand{\bef}{\begin{figure}[htbp]\begin{center}}
\newcommand{\eef}{\end{center}\end{figure}}
\newcommand\Perimeter{Perimeter Institute for Theoretical Physics, Waterloo, ON N2L 2Y5, Canada}

\begin{document}

\title{Searching for Light Dark Matter with the SLAC Millicharge Experiment}
\author{M.~Diamond}          \thanks{miriam.diamond@alumni.carleton.ca}  \affiliation{\Perimeter} 
\author{P.~Schuster}           \thanks{pschuster@perimeterinstitute.ca}  \affiliation{\Perimeter} 

\date{\today}
\begin{abstract}
New sub-GeV gauge forces (``dark photons'') that kinetically mix with the photon provide a promising scenario for $\MeV - \GeV$ dark matter, 
and are the subject of a program of searches at fixed-target and collider facilities around the world.  
In such models, dark photons produced in collisions may decay invisibly into dark matter states, thereby evading current searches. 
We re-examine results of the SLAC mQ electron beam dump experiment designed to search for millicharged particles, and find 
that it was strongly sensitive to any secondary beam of dark matter produced by electron-nucleus collisions in the target. 
The constraints are competitive for dark photon masses in the $\sim 1-30 \ \MeV$ range, covering part of the parameter space
 that can reconcile the apparent $(g-2)_{\mu}$ anomaly. Simple adjustments to the original SLAC search for millicharges may 
 extend sensitivity to cover a sizeable portion of the remaining $(g-2)_{\mu}$ anomaly-motivated region.
 The mQ sensitivity is therefore complementary to on-going searches for visible decays of dark photons. 
 Compared to existing direct detection searches, mQ sensitivity to electron-dark matter scattering cross sections
  is more than an order of magnitude better for a significant range of masses and couplings in simple models. 
\end{abstract}


\maketitle

Identifying dark matter is one of the most pressing open problems in fundamental physics. 
Although a rich experimental program continues to probe dark matter (DM) interactions for masses in the $10 \ \GeV$- $\TeV$ range, 
sensitivity to DM at lower masses remains remarkably poor. 
There are well motivated scenarios of sub-$\GeV$ DM, especially those that include new gauge forces (``dark forces") that 
kinetically mix with the photon --  these models can account for the observed relic density consistently with 
all available data, and have been the focus of intense discussion in the literature \cite{Holdom:1985ag,*ArkaniHamed:2008qn,*Pospelov:2007mp}.

In this note, we show that the electron beam dump millicharge search at SLAC (mQ) was sensitive to sub-GeV DM 
interacting through dark photons. 
In a simple model, we compute the total detection yield for $\MeV$-scale DM components 
that would have been produced in the mQ target. 
We use these yields and the original mQ analysis to establish constraints on such DM.
The new constraints cover part of the parameter space that can reconcile the apparent $(g-2)_{\mu}$ anomaly,
and future adjustments to the original analysis may significantly extend sensitivity. 
We also provide estimates for the level of sensitivity that might be attained with a re-designed version of this experiment at 
modern high-intensity electron beam facilities. 
These results highlight the potential for using electron beam dump experiments to powerfully probe {\it any} DM components
(or other long-lived particles) that couple to leptons and quarks (see \cite{NewPaper}), and they complement the on-going effort 
to search for dark photons in visible decay channels \cite{Abrahamyan:2011gv,*Merkel:2011ze,*Essig:2010xa,*Aubert:2009cp,*Echenard:2012hq,*Babusci:2012cr,*Adlarson:2013eza}.

As a simple example, we consider a benchmark model consisting of a long-lived fermion $\chi$ coupled to a 
dark sector $U(1)_D$ gauge boson that kinetically mixes with the photon. The Lagrangian is
\be
\mathcal{L} \supset \epsilon _Y F^{Y, \mu \nu} F _{D,\mu \nu} - \frac{1}{4} F_D^{\mu \nu} F _{D,\mu \nu} + \frac{m_{A'}^2}{2} A'^{\mu} A' _{\mu} + g_D J ^\mu _{A'} A' _\mu \nonumber
\ee
where $F_{Y,\mu \nu} = \partial _{[ \mu} B _{\nu ]}$ is the field strength tensor for Standard Model (SM) hypercharge $U(1)_Y$, $F_{D,\mu \nu} = \partial _{[ \mu} A' _{\nu ]}$ for the new $U(1)_D$, and $J ^\mu _{A'}$ is the interaction current of the $A'$ with any dark-sector fields, in this case a fermion $\chi$.  
We define $\epsilon \equiv \epsilon _Y \cos \theta _W$ where $\theta _W$ is the SM weak mixing angle, and $\epsilon ^2 = \frac{\alpha_{dark}}{\alpha_{EM}} $. A field redefinition removes the kinetic mixing term and generates a coupling $\epsilon e A' _\mu J ^\mu _{EM}$ between the $A'$ and SM electrically-charged particles.  This effectively gives charged particles a small dark force charge, without giving dark sector particles electric charge.  Kinetic mixing with $\epsilon \sim 10^{-3} - 10^{-2}$ can be generated by loops of heavy fields charged under both $U(1)_D$ and U(1)$_Y$, and is a natural range to consider \cite{fixed-target-exp,*dark-decays}.

Previous literature has considered numerous constraints on sub-GeV DM derived from the CMB, supernovae, B-factory searches, rare Kaon decay measurements, 
and precision $(g-2) _\mu$ and $(g-2) _e$ measurements \cite{int-frontier}. 
For comparison to the mQ sensitivity, we include the constraints relevant for the low $m_{A'}$ range. 
A companion paper \cite{NewPaper} discusses the viability of using the simple benchmark Lagrangian above to model fixed-target physics, where $\chi$ can be all of or a sub-dominant part of 
the DM consistent with all available data. 

In the mQ experiment, 1.35 Coulombs ($8.4\times 10^{18} e^-$) of $29.5 \ \GeV$ electrons were deposited on a tungsten production target. 
Approximately 90~m of sandstone separated the target from the detector (Bicron-408 plastic scintillator), which was sensitive to signals as small as a single scintillation photon and subtended a solid angle of $2 \times 2~$mrad$^2$.  
SM particles essentially ranged out in the sandstone, while any penetrating particles like mQ's were able
to reach the detector and trigger a small scintillation signal \cite{slac-mQ}. 
Collected data consisted entirely of timing and height of PMT pulses.  No significant signal was found over a rather large ($\sim 146 $K) but well-measured instrumental background \cite{SLAC-thesis}. 

As illustrated in Figure~\ref{fig:mQ-setup}, this setup would have produced significant numbers of $A'$s in the target via a bremsstrahlung-like process.  
We examine the case of prompt invisible decay $A' \rightarrow \chi \bar{\chi}$; the $\chi$'s would have traversed the sandstone given their large 
mean free path. The secondary beam of $\chi$'s could have deposited energy in the mQ detector via $Z^2$-enhanced elastic scattering
off carbon nuclei (and sub-dominantly though quasi-elastic $\chi$-nucleon scattering, which we neglect).
\bef
\includegraphics [trim = 20mm 10mm 25mm 50mm, clip, width = 0.48\textwidth]{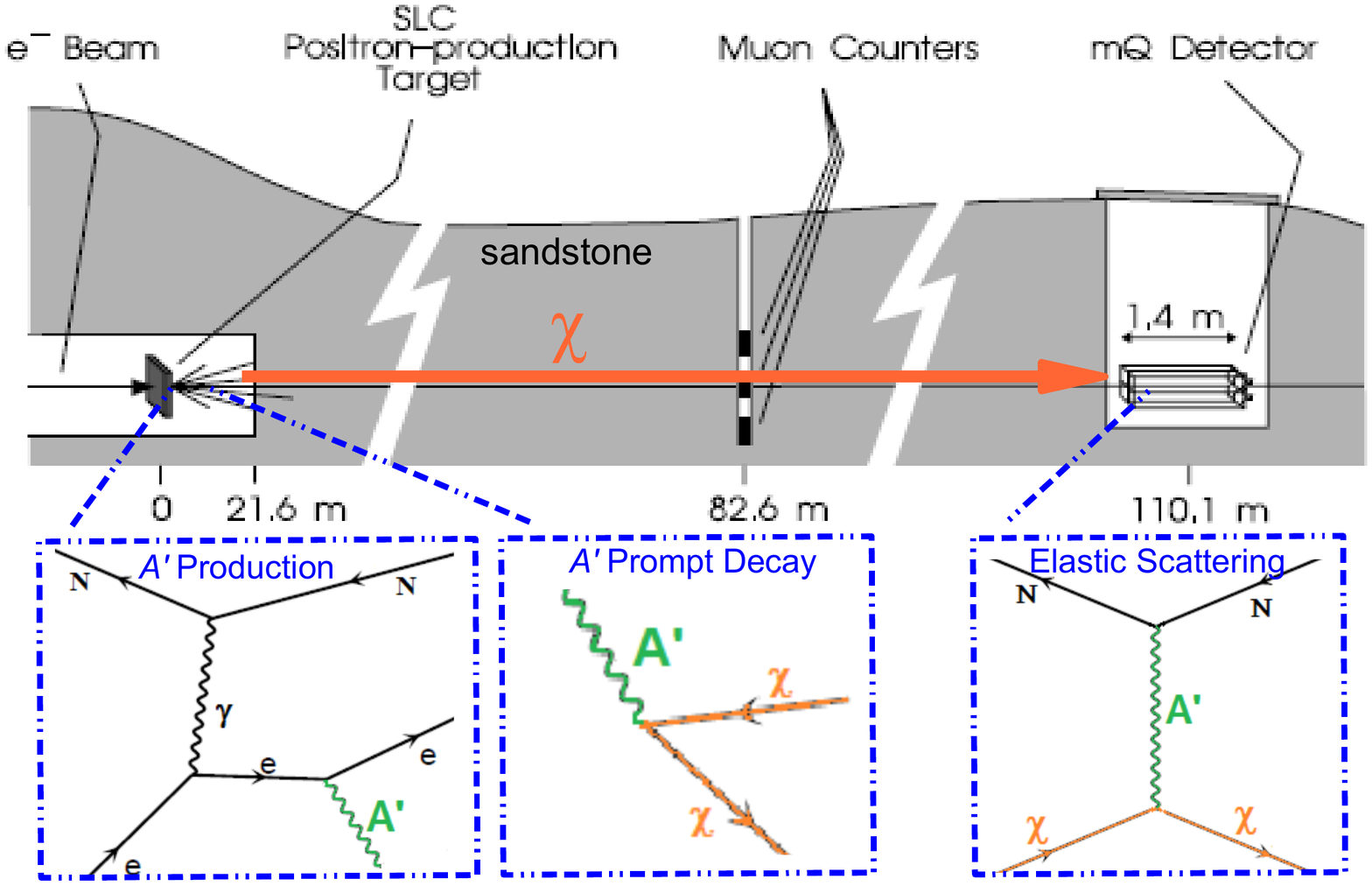}
\caption{Layout of the SLAC mQ experiment \cite{slac-mQ}.  We investigate the possibility of $A'$ production in the target, followed by prompt decay to long-lived dark sector particles $\chi$, which could traverse the sandstone and undergo elastic scattering off carbon nuclei in the detector.}
\label{fig:mQ-setup}
\eef

Our analysis assumes $2 m_{\chi} < m_{A'}$ (on-shell $A'$), but we expect this approach to have sensitivity 
even for $2 m_{\chi} > m_{A'}$ where $\chi$'s are produced via an \emph{off-shell} $A'$ (see \cite{fixed-target-exp,*dark-decays}). 
We used the procedure in \cite{fixed-target-exp,*dark-decays}, based on a variation on the Weizsacker-Williams method, for computing $A'$ production. 
We also simulated all reactions using MadGraph and MadEvent 4 \cite{madgraph}. 
We assigned to the $e^- e^- A'$ vertex the coupling $e_{EM} \epsilon$, and to the $N_t N_t \gamma$ vertex $e_{EM} Z_t \eta $ (for target nucleus $N_t$ of atomic number $Z_t$, with $\eta _t$ the square root of the form-factor in \cite{fixed-target-exp,*dark-decays}.)

The typical emission angle for the $A'$ relative to the beam is parametrically smaller than the 
opening angle of $A'$ decay products, and is collinear to a good approximation. 
Neglecting $m_e$, 
\be \label{eqn:xs-aprod}
\frac{d \sigma _{A' prod}}{d x} \approx \frac{8 Z^2 ~\alpha^3 ~\epsilon ^2 ~x \times Log}{3 m_{A'} ^2} \left( 3 + \frac{x^2}{1-x} \right) ,
\ee
where $Z$ is the atomic number of the target nucleus, $x\equiv E_{A'}/E_0$ with $E_0$ the lab-frame energy of the beam electron, 
and $Log$ is an $O(10)$ factor dependent upon kinematics, atomic screening, and nuclear size effects \cite{fixed-target-exp,*dark-decays}.

Since the angular size of the mQ detector was $\theta_d\approx 2 $ mrad, angular acceptance limits overall sensitivity. Produced $A'$s typically carry most of the beam energy, with $x_{median} \sim 1 - \min \left( \frac{m_e}{m_{A'}}, \frac{m_{A'}}{E_0} \right)$ \cite{fixed-target-exp,*dark-decays}.  In the $A' \rightarrow \chi \chi$ decay, the angle $\theta_{\chi}$ of the $\chi$'s relative to the beamline scales as $\frac{m_{A'}}{E_0}$. The angular distribution of $\chi$ is shown in Figure \ref{fig:ang-distrib} for reference. 
\bef
\includegraphics[trim = 10mm 0mm 2mm 0mm, clip, width=0.48\textwidth]{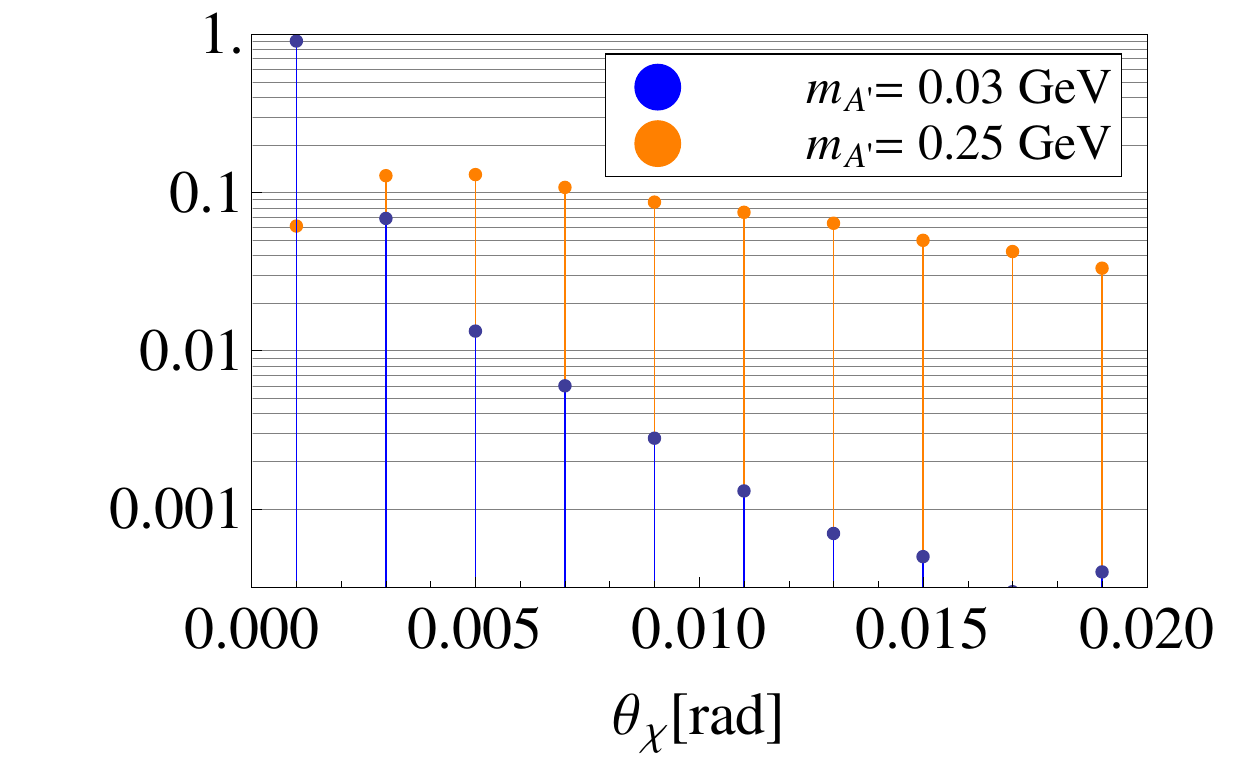}
\caption{Sample $\chi$ angular distributions, generated using MG/ME ($m_{\chi} = 10~$MeV).  Normalized to 1.  Angular acceptance of mQ detector is 0.002~rad.  $m_{A'}=0.03$~GeV (steep curve): 90\% accepted.   $m_{A'}=0.25$~GeV (shallow): 6\% accepted.}
\label{fig:ang-distrib}
\eef

For the coherent scattering illustrated in Figure \ref{fig:mQ-setup}, we assigned the $N_d N_d A'$ vertex the coupling $e_{EM} Z_d \eta _d \epsilon$ (for detector nucleus $N_d$ carbon.)  
With $T$ the lab-frame kinetic energy of the recoiling nucleus of mass $M \ll T$, the coherent scattering cross-section (neglecting $\eta$) is approximately
\be \label{eqn:xs-scatter}
\frac{d \sigma_{\chi N scatt}}{dT} \approx \frac{-8 \pi \alpha \alpha _D \epsilon ^2 Z^2 M}{(m_{A'}^2 + 2 M T)^2}.
\ee
The recoil distributions in full simulation for representative $m_{A'}$ are shown in Figure \ref{fig:nuc-recoil}.
The nuclear recoil energy is typically $O(0.025-1.0)$~MeV.  
Based on neutron scattering experiments with plastic scintillator, a proton recoiling with kinetic energy 1~MeV should produce $\sim 59$ PEs, and 0.1~MeV $\sim 2.3$ PEs.
The quenching factor for a C nucleus is about half that for a proton \cite{scatt-physrev,*Harvey-Hill,*nucl-ex-scint}. Therefore a 1~MeV recoiling C should produce $\sim 30$ PEs, and 0.1~MeV $\sim 1$ PE.  Figure \ref{fig:nuc-recoil} shows that with a $\sim 0.1$~MeV threshold for producing a PE, about 20\% of the $\chi$-C events would produce at least a single PE at $m_{A'} = 0.03~$GeV, and 90\% at $m_{A'} = 0.25~$GeV. 

\bef
\includegraphics[trim = 10mm 0mm 2mm 0mm, clip, width = 0.48\textwidth]{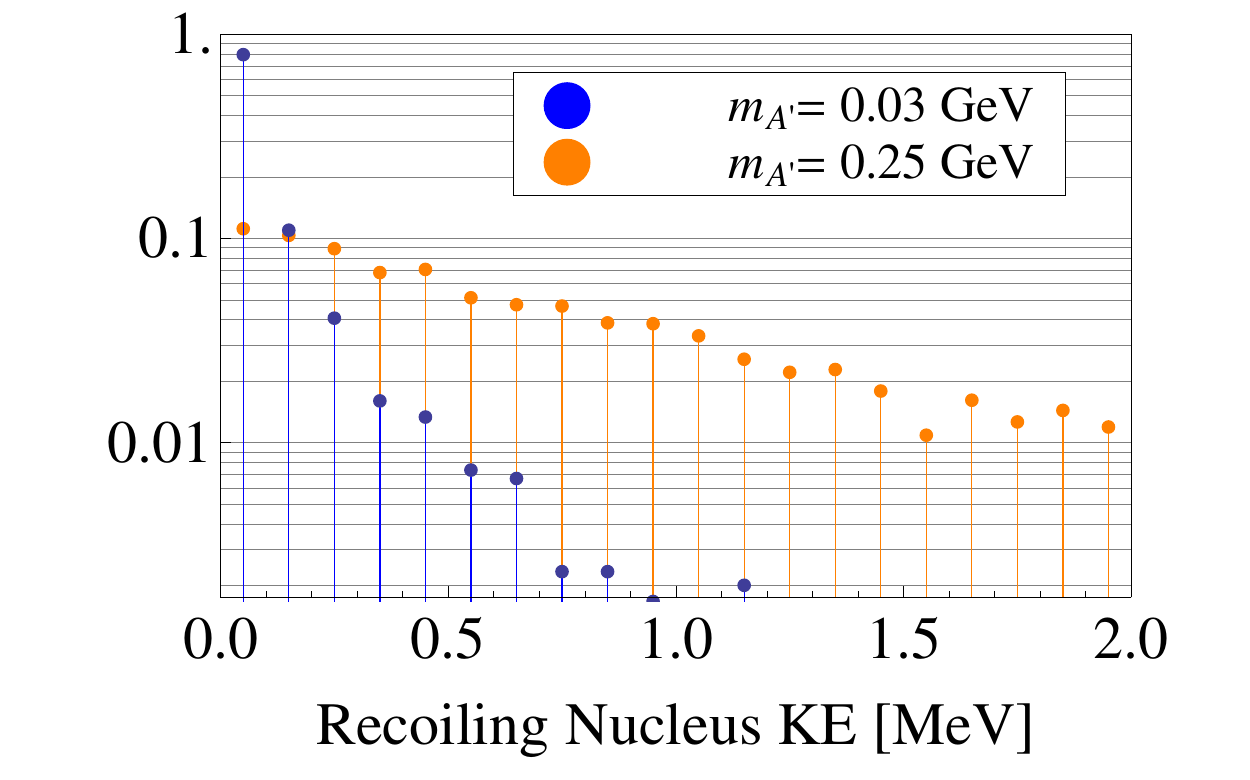}
\caption{Sample nuclear recoil distributions, generated using MG/ME ($m_{\chi} = 10~$MeV), at the approximate median $\chi$ lab-frame energy (12 GeV for $m_{A'} = 0.03$~GeV, 27 GeV for $m_{A'} = 0.25$~GeV.)  Normalized to 1.}
\label{fig:nuc-recoil}
\eef

In finding the total number of $\chi$ produced in the target, we can neglect $\chi$ production in lower energy showers initiated by the beam electron 
because the angular acceptance of mQ is small. 
To account for the more important effect of the energy loss of the beam $e^-$ as it traverses the target, we use an ``effective" radiation length of 
$T_{eff} =1$. This can be justified as follows. 
For the small angular size of mQ, the angular acceptance scales as $E^2$ (for low $A'$ masses), where $E$ is the beam electron energy. 
Thus, the $E^2$-weighted average of the beam energy distribution integrated over the thickness of the target (6 radiation lengths) and energy
yields an effective thickness (in units of radiation length). 
Using the beam energy distribution $I(E,E_0)$ in \cite{fixed-target-exp,*dark-decays}, we obtain 
$T_{eff} =\int ds \int dE (E/E_0)^2 I(E,E_0,s) = \frac{3}{2}\ln{2}\approx 1$.
To a good approximation, the differential production yield for fixed $\chi$ energy $E_{\chi}$ is
\be
\frac{dN_\chi}{dE_\chi} \approx 2 T_{eff} \frac{N_e N_0 X_0}{A} ~\frac{d\sigma_{\chi prod}}{dE _\chi}
\ee
where $N_e$ is the total number of beam $e^-$ incident on target, $N_0$ is Avogadro's number, $X_0$ is the unit radiation length of target material, and $A$ is the target atomic mass.
The differential production cross section at fixed $\chi$ energy, $\frac{d\sigma_{\chi prod}}{dE _\chi}$, was computed with full simulation.  
To find the number of expected $\chi-N_d$ scattering events in the mQ detector, $N_{evts}$,
we include angular acceptance cuts with full simulation, which reduces $N_\chi$ to $N_{\chi ~ acc}$.
The final yield is then
\be \label{eqn:Nchi}
N_{evts} = \int dE_\chi ~\frac{dN_{\chi ~ acc}}{dE_\chi} ~ \sigma _{\chi N scatt}(E_\chi) ~ t_{Det} ~ \rho_{Det},
\ee
where $t_{Det}$ is the detector thickness, and $\rho_{Det}$ is the number density of C nuclei in the detector.

An order-of-magnitude estimate can be obtained by $N_{\chi ~acc}~\approx~2 N_e P_{prod} P_{scatt} F$, where the probability per beam $e^-$ to produce a $\chi$ pair is 
$$P_{prod} \approx 1.2 \times 10^{-11} \left(\frac{\epsilon ^2}{10^{-6}}\right) \left(\frac{0.05 \GeV}{m_{A'}}\right)^2,$$ 
the probability of $\chi$-C coherent nuclear scattering is 
$$P_{scatt} \approx 2.5 \times 10^{-8} \left(\frac{\epsilon ^2}{10^{-6}}\right) \left(\frac{0.05 \GeV}{m_{A'}}\right) \frac{\alpha _D}{\alpha _{EM}},$$ 
and $F\sim (\theta_d E_0/m_{A'})^2$ is the fraction of $\chi$'s that pass angular acceptance cuts.  Table \ref{tab:num-evts} gives the simulated cross-sections and production totals, along with the corresponding analytical estimates, for one example set of parameter values -- the agreement is quite good. 

\begin{table}[h]
\begin{center}
  \begin{tabular}{ | l || l | l |}
    \hline
  	Quantity & Simulated Value & Analytic Estimate \\
    \hline
    $\sigma _{\chi prod}$ [pb] & 37722 & 38000 \\
	No. $\chi$ produced & $1.39 \times 10^{10}$ & $1.42 \times 10^{10}$ \\
	$F$ & 0.52 &  0.60\\
	$\sigma _{\chi Nscatt,rel}$ [pb] & 3950 &  4200\\
	No. scattering evts & 189 & 239 \\
    \hline
  \end{tabular}
\end{center}
\caption{For the benchmark point $m_{A'} = 0.05$~GeV, $m_\chi = 0.01$~GeV with $\epsilon = 10^{-3}$, $\alpha _D = \alpha _{EM}$: $\chi$ production cross-section, total no. $\chi$ produced, fraction of $\chi$ that pass the angular acceptance cut, $\chi$-C coherent scattering cross-section for relativistic $\chi$, and total no. of $\chi$-C coherent scattering events in the mQ detector.}
\label{tab:num-evts}
\end{table}

Using five ``benchmark" points with $m_\chi = 0.01~$GeV, in the $m_{A'}= 0.03 - 0.25 \ \GeV$ range, we evaluated the 
limits in the $(m_{A'}, \epsilon)$ parameter space by comparing total yields to single PE mQ background measurements. 
The mQ data analysis estimated $\sim 94\%$ of the 146061 background events involved only a single PE \cite{SLAC-thesis}.  
For $m_{A'}> 100 \ \MeV$, scattering events should produce much more than one PE, so it should be possible to use a PMT pulse-height cut to help separate $\chi$ signal from background.  
It is reasonable to expect such a cut to improve S/B by at least an order of magnitude in the higher $m_{A'}$ range because the vast majority of the background is single PE noise. 
Figure \ref{fig:param-space-nuc} shows the $2 \sigma$ constraints that would be obtained for $m_\chi = 10~$MeV with no background reduction, and with $100 \times$ the reported S/B.  
Note, Figure \ref{fig:param-space-nuc} assumes every scattering event in the detector produces at least one photo-electron and is observed.  
Losses from failure to produce any PEs could reduce sensitivity by a factor of $\sim 2$ in the lowest $(\sim 30 \ \MeV)$ part of the $m_{A'}$ range. 

Given significant background reduction, mQ would be able to cover a sizeable swath of unexplored parameter space, including part of
the $(g-2) _\mu$ anomaly-motivated region for $m_{A'}\sim 0.03 - 0.160$~GeV. 
It should be noted that there is currently a MiniBooNE proposal for further running specifically to cover this range \cite{miniboone}. Likewise, LSND 
could likely impose constraints at the level of $\epsilon \sim 10^{-4}-10^{-3}$ for $m_{A'} < O$($100 \ \MeV$), $m_{\chi} \ll \frac{m_{A'}}{2}$ though 
no analysis is yet publicly available. 

Our analysis results can be interpreted as constraints on electron-$\chi$ scattering cross sections $\sigma_{e}$, which can also 
be probed by direct detection. 
Recent results from XENON10 established limits on $\sigma _e$ as a function of DM mass in the 1--1000 MeV range \cite{xenon10}.  
Using ``benchmark" points shown in Figure \ref{fig:param-space-dm}, 
we employed mQ constraints on $(m_{A'}, \epsilon)$ to establish constraints on $\sigma _e$ via $ \sigma_e = \frac{16 \pi \alpha _{EM} \alpha _D \epsilon ^2 m_e^2}{m_{A'}^4}$.  If $\chi$ accounts for \emph{all} the DM, mQ sets limits more stringent than XENON10 for $m_\chi < 20$~MeV.  $\chi$ could instead be a sub-dominant DM component, in which case XENON10 constraints are weakened.

\bef
\includegraphics [width = 0.48\textwidth]{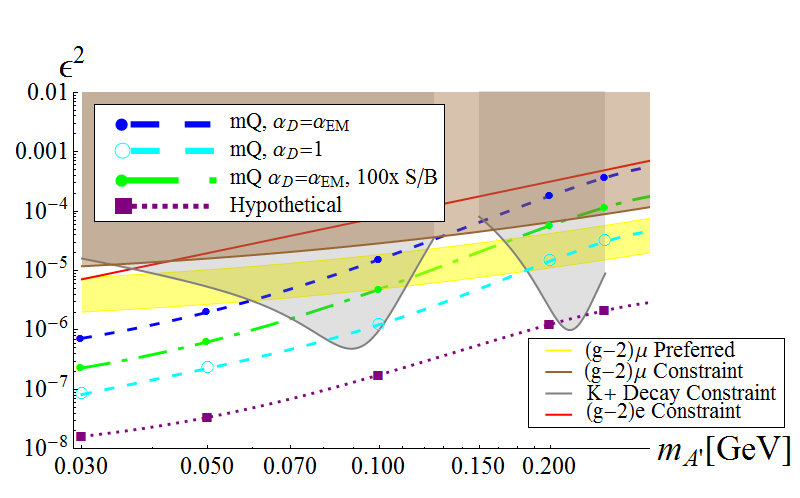}
\caption{For each benchmark $m_{A'}$ with $m_\chi=0.010~$GeV, the $\epsilon$ that would correspond to a $2 \sigma$ result in SLAC mQ.  Note the dependence on $\alpha _D$, and the improvement that would come from achieving $100 \times$ the reported mQ S/B.  These results change fairly little with $m_\chi$.  Overlaid on existing $A' \rightarrow inv$ constraints; $(g-2)_e$ is a $2 \sigma$ power constrained limit \cite{NewPaper}.  Yellow band: $(g-2) _\mu$ anomaly-favoured \cite{g-2e,*Pospelov:2008zw}.  Note, LSND would be expected to provide additional constraints, at the level of $\epsilon ^2 \sim 10^{-6} - 10^{-8}$ for $m_{A'} \leqslant 0.05$~GeV.}
\label{fig:param-space-nuc}
\eef

\bef
\includegraphics [trim = 5mm 0mm 5mm 0mm, clip, width = 0.48\textwidth]{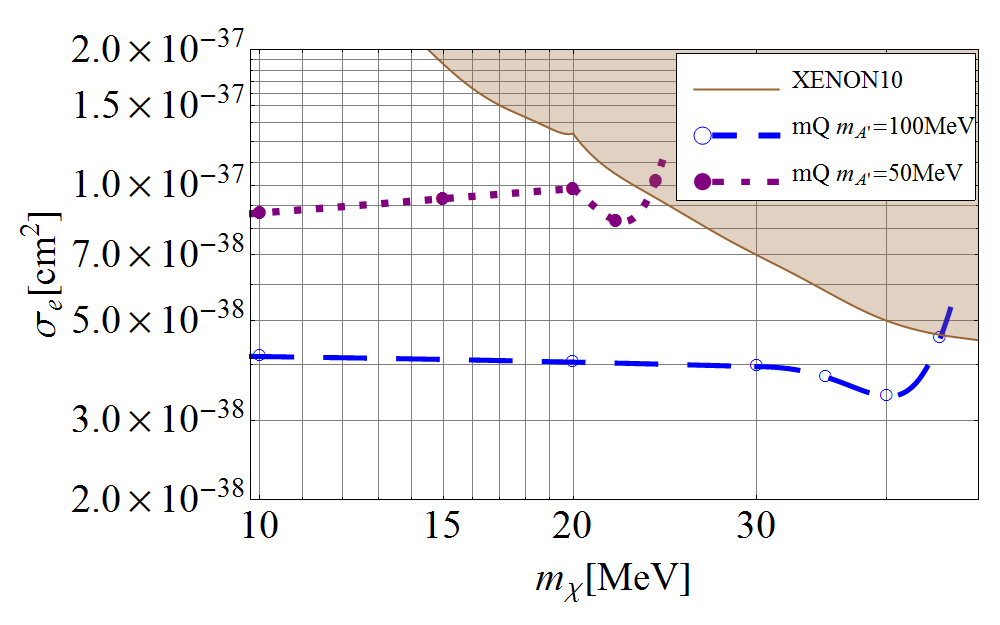}
\caption{For benchmark $m_\chi$ with $m_{A'}=0.10~$GeV and $m_{A'}=0.05~$GeV ($\alpha _D=\alpha _{EM}$), the constraints on scattering cross-section of DM off $e^-$ corresponding to a $2 \sigma$ result in mQ, assuming the reported mQ signal-to-background.  Overlaid on the XENON10 direct detection results  \cite{xenon10}; direct comparison valid only assuming $\chi$ accounts for all the DM in the universe. }
\label{fig:param-space-dm}
\eef

It is convenient to consider mQ because the data already exists -- but this experiment was not optimized for light DM searches.  Characteristics that would make future $e^-$ beam-dumps even more effective for this purpose include optimal sensitivity to quasi-elastic $\chi$-nucleon processes, broader angular acceptance, greater luminosity, and an effective background-rejection scheme \cite{NewPaper}.  The main backgrounds are typically intrinsic detector noise, cosmic rays, $\gamma$'s from ambient radioactivity, and fast neutrons (produced from the target). Neutral-current $\nu$ interactions are negligible \cite{SLAC-thesis}.  
As an exercise, each benchmark point in Figure \ref{fig:param-space-nuc} was re-calculated for a luminosity of $10^{22}$ electrons, with no angular acceptance cuts.  
This luminosity could be reasonably achieved at a facility such as Jefferson Laboratory or a future Linear Collider. 
Sensitivity to 500 signal events for example (realistic for $>>1$ PE yield signals), would cover an impressive swath of parameter space (dotted line in Figure \ref{fig:param-space-nuc}). 

In conclusion, we find the SLAC mQ search is indeed relevant for exploring the parameter of models where a 
dark photon of mass $\sim 30-300 \ \MeV$ decays to lighter, long-lived $\chi$'s.  
This includes a parameter region in which dark photon models can alleviate the current $(g-2) _\mu$ discrepancy, 
and adjustments to the original SLAC analysis are expected to strengthen the constraints -- or make a discovery -- in this region.  
In a broader context, our analysis provides a proof-of-concept for the use of $e^-$ beam-dumps to search for DM particles with masses 
of tens to hundreds of MeV, a regime that poses great difficulty for direct detection and collider experiments.  
In simple models, we find that mQ constrains the DM-electron scattering cross-section 
$\sigma_e \lesssim 10^{-38} - 10^{-37}$~cm$^2$ for $m_\chi \sim 10-40 \ \MeV$ -- up to an order of magnitude stronger than the leading direct-detection limits where applicable. 

\begin{acknowledgments}
{\it Acknowledgments} We thank Natalia Toro, Gordan Krnjaic, and Eder Izaguirre for helpful feedback and discussion. This research was supported in part by Perimeter Institute for Theoretical Physics. Research at Perimeter Institute is supported by the Government of Canada through Industry Canada and by the Province of Ontario through the Ministry of Research and Innovation.
\end{acknowledgments}

\bibliographystyle{apsrevM}
\bibliography{mQprl}
\end{document}